\documentclass[cameraready]{Interspeech}
\usepackage{cite}
\usepackage{amsmath,amssymb,amsfonts}
\usepackage{algorithmic}
\usepackage{graphicx}
\usepackage{textcomp}
\usepackage{amsmath,bm}
\usepackage{multirow}
\usepackage{multicol}
\usepackage{xcolor}
\usepackage{verbatim}
\usepackage{arydshln}
\usepackage{adjustbox}

\title{EmoEUS: Uncertainty Supervision for Multimodal Emotion Recognition in Conversation}

\author[affiliation={1},orcid=0009-0003-0760-8350]{Zilong}{Huang}
\author[affiliation={1},orcid=0000-0001-9133-3000]{Kong Aik}{Lee}
\author[affiliation={1},orcid=0009-0007-9654-9223]{Junjie}{Li}
\author[affiliation={2},orcid=0000-0002-0519-7434]{Zhe}{Li}
\author[affiliation={1},orcid=0000-0001-8854-3760]{Man-Wai}{Mak}

\address{ 
    $^1$ Dept. of Electrical and Electronic Engineering, The Hong Kong Polytechnic University \\
    $^2$ Speech, Language, and Cognition Laboratory, The University of Hong Kong
}

\email{\{zi-long.huang, kong-aik.lee, junjie98.li, man.wai.mak\}@connect.polyu.hk, zheli8@hku.hk}


\usepackage{comment}

\begin{document}

\maketitle

\begin{abstract}
Multimodal emotion recognition in conversation (MERC) can leverage multimodal and contextual cues to boost recognition performance. However, existing fusion approaches in MERC often ignore modality-specific uncertainty across utterances caused by conflicting cues, varying noise, and missing modality-specific signals. We propose EmoEUS, an explicit uncertainty supervision framework for MERC. EmoEUS performs uncertainty-aware multimodal fusion by dynamically weighting modalities using learned variance estimates. We also introduce an explicitly supervised loss that aligns each utterance's predicted variance with the distance between the utterance's distributional representation and its emotion- and modality-specific cluster center. Experiments on IEMOCAP and MELD show that EmoEUS consistently outperforms state-of-the-art methods.

\end{abstract}

\section{Introduction}
\label{sec:intro}

Multimodal \textbf{e}motion \textbf{r}ecognition in \textbf{c}onversation (MERC) is important for applications such as human-computer interaction, intelligent medical care, and affective computing \cite{poria2019emotion}. MERC analyzes dialogue turns from multiple modalities (e.g., text, audio, and visual) and assigns one emotion label per utterance \cite{poria2019emotion,mm-nodeformer}. Prior work focuses on integrating multimodal information, modeling cross-modal interactions, and improving contextual feature representations \cite{glodek2011multiple,shi-huang-2023-multiemo,extraren2025bav}.

Current MERC approaches predominantly adopt Transformer-based architectures to model multimodal conversational contexts \cite{chudasama2022m2fnet,li2023cfn}. By jointly modeling multimodal input sequences, these methods effectively capture long-range contextual dependencies and cross-modal interactions, producing powerful fused representations for emotion recognition \cite{mm-nodeformer}. Recent studies further enhance these architectures through improved distillation mechanisms, graph-based modeling, or modality-specific encoders to better capture contextual cues and speaker dependencies in conversations\cite{extraidir,disstilmerc,extradis}.
Despite the advances in representation learning \cite{extraunet,extradenoising}, multimodal conversational data inherently exhibits significant uncertainty, which remains insufficiently addressed by existing fusion models \cite{wu2014survey,SDT,uncertain1,uncertain2}. While uncertainty modeling has been widely studied in multimodal learning to assess input reliability and improve model robustness \cite{li2025disentangling,cv1,cv2,MAP}, its exploration in MERC is still limited. Most existing MERC methods implicitly assume all modalities are uniformly reliable across every utterance, overlooking the substantial variations in their quality and informativeness \cite{uncertain1}. In practice, modality quality and emotional contribution vary substantially due to noise, missing cues, or conflicting signals, making uniform fusion suboptimal \cite{uncertain2}. 

Although recent works have found the importance of introducing modality-specific uncertainty for MERC \cite{meng2024masked,COLD}, they typically model uncertainty implicitly through classification loss alone, lacking explicit supervision to dynamically leverage the emotional information of each modality for adaptive fusion.
To address these limitations, we propose \textbf{e}xplicit \textbf{u}ncertainty \textbf{s}upervision for multimodal \textbf{emo}tion recognition in conversation (\textbf{EmoEUS}). EmoEUS transforms point-wise features into distributional representations with variance via a context-level distribution estimator module (ContextDEM) and an explicitly supervised loss (ESL), and it performs fusion through an uncertainty-aware multimodal fusion (UAMF) mechanism. This design quantifies modality reliability and enables adaptive multimodal fusion under uncertain conversational contexts. Our main contributions are as follows:

\begin{itemize}
\item We propose a context-level distribution estimator module (ContextDEM) that models utterance-level uncertainty by transforming multimodal features into distributional representations with means and variances.

\item We introduce an uncertainty-aware multimodal fusion (UAMF) mechanism that dynamically adjusts modality contributions according to their estimated uncertainty.

\item We design an explicitly supervised loss (ESL) based on distributional cluster centers and 2-Wasserstein distance to provide explicit supervision for uncertainty estimation.

\end{itemize}

\begin{figure*}[t]
\centering
\includegraphics[width=1.0\textwidth]{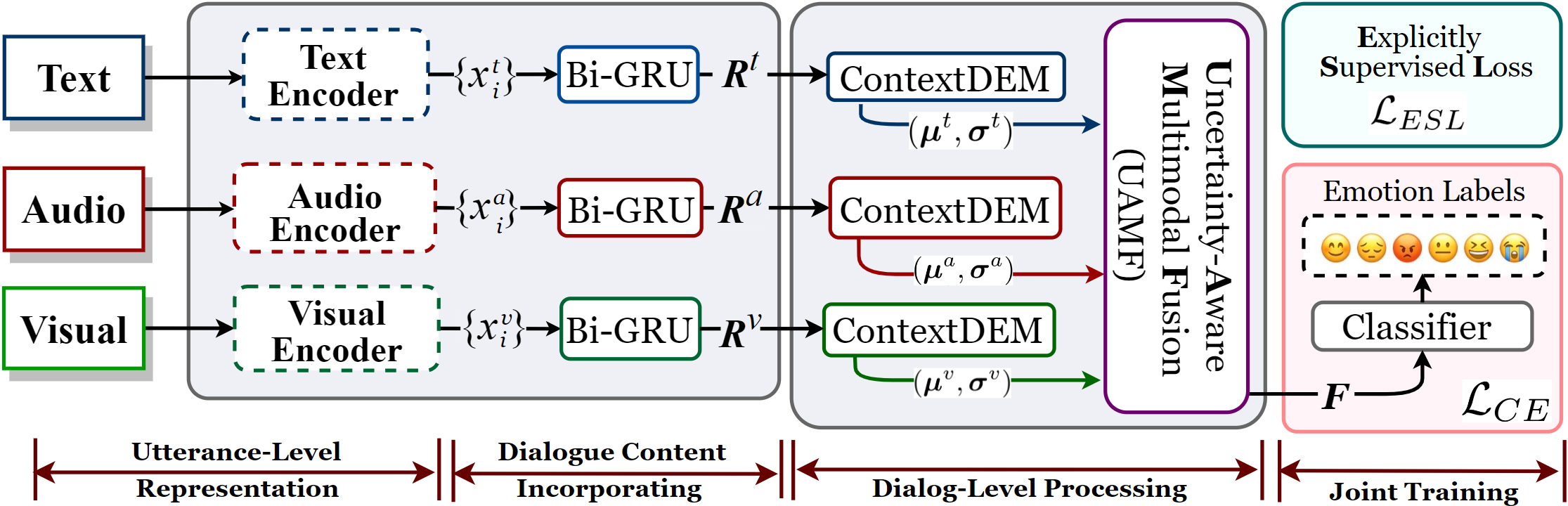}
\caption{The overall architecture of the EmoEUS framework. EmoEUS enhances MERC performance through explicit uncertainty supervision and uncertainty-aware multimodal fusion. Solid and dashed rectangles denote trainable and frozen models, respectively. }
\label{overall}
\vspace{-1.2em}
\end{figure*}

\section{Methodology}
\subsection{Task Definition}
Consider a dialogue consisting of $k$ temporally ordered utterances, denoted as ${\cal U}=\{u_i\}_{i=1}^k$, where individual utterances have corresponding emotion labels $\{y_i\}_{i=1}^k$ and speaker labels $\{s_i\}_{i=1}^k$. The emotion label ${y_i}$ belongs to a predefined set of emotions ${\cal Y}$. Each utterance ${u_i}$ is accompanied by multimodal data, including a video clip, an audio segment, and a text transcript. Formally, we represent a dialogue as follows:
\begin{equation}
    {\cal U}\times{\cal Y}\times{\cal S}  = \left\{ \{ u_i^{(\delta)} ,{y_i},{s_i}\} \ | \ \delta  \in \{ a,t,v\} ,{y_i}\in{\cal Y}, {s_i} \in {\cal S} \right\}, \label{eq1}
\end{equation}
where $u_i^{(\delta)}$ denotes the modality-specific representation (text, audio, or video) of the $i$-th utterance. The ERC task involves predicting the emotion label ${\hat y_i}$ for each utterance ${u_i}$ in the dialogue, leveraging information from all $k$ utterances.

\subsection{Overall Architecture}

As shown in Fig.~\ref{overall}, EmoEUS consists of three components. First, ContextDEM maps multimodal features to Gaussian distributions \((\boldsymbol{\mu}, \boldsymbol{\sigma}^2)\) to model modality-specific uncertainty. Second, an uncertainty-aware multimodal fusion (UAMF) module integrates these distributions by weighting modalities based on their uncertainties. Finally, the explicitly supervised loss $\mathcal{L}_{ESL}$ aligns the predicted variance with the 2-Wasserstein distance between the emotion-specific cluster center and the utterance's distribution representation, providing supervision for uncertainty estimation. The model is optimized using the classification loss $\mathcal{L}_{CE}$ together with the weighted $\mathcal{L}_{ESL}$.

\subsection{Contextual Feature Extraction}
For each dialogue utterance $u_i$ ($i = 1,\ldots,k$) with textual ($u_i^t$), acoustic ($u_i^a$), and visual ($u_i^v$) modalities, we employ modality-specific pre-trained encoders to extract features:
\begin{equation}
    \bm{x}_i^\delta = \operatorname{ModalityEncoder}(u_i^\delta), \text{ }\delta \in \{t,a,v\},
\end{equation}
where the features are obtained by average pooling the final layer's outputs of individual encoders. To capture sequential dependencies, the features are processed by bidirectional GRUs:
\begin{equation}
    \bm{R}^\delta = \operatorname{Bi-GRU}(\bm{x}_1^\delta, \ldots, \bm{x}_k^\delta) \in \mathbb{R}^{D_r \times k},
\end{equation}
where $D_r$ is twice the hidden state dimension of the GRUs.

\subsection{Context-level Distribution Estimation}
\begin{figure}[t]
\centering
\includegraphics[width=0.47\textwidth]{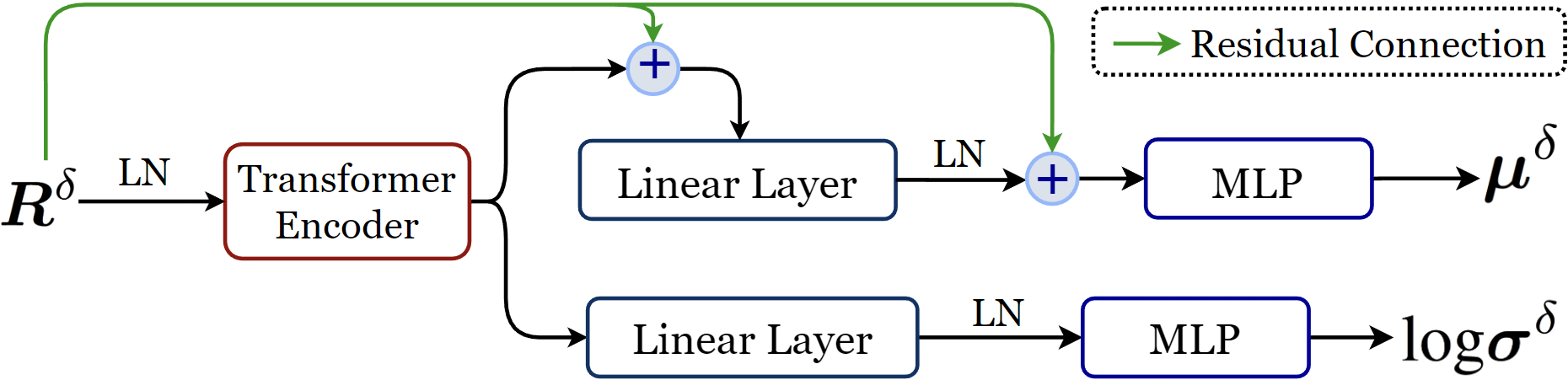}
\caption{Framework of ContextDEM, where LN and MLP denote LayerNorm and multi-layer perceptron, respectively.}
\label{contextpde}
\vspace{-1.75em}
\end{figure}
To model semantic uncertainty, we propose a context-level distribution estimator module (ContextDEM), as shown in Fig.~\ref{contextpde}. Given modality-specific features $\bm{R}^\delta$, the ContextDEM first enhances global contextual interactions via a Transformer encoder, followed by residual connections to preserve original context. The resulting features are then fed into two parallel MLP branches to independently predict the mean $\boldsymbol{\mu}^\delta$ and log-variance $\log \boldsymbol{\sigma}^\delta$ of a Gaussian distribution:\footnote{For notational simplicity, we denote $\bm{\sigma}$ as the variance instead of the usual $\bm{\sigma^\text{2}}$.}
\begin{equation}
    (\boldsymbol{\mu}^\delta, \log \boldsymbol{\sigma}^\delta) = \operatorname{ContextDEM}(\bm{R}^\delta), \text{ } \delta \in \{t,a,v\},
\end{equation}
where $\boldsymbol{\mu}^\delta, \boldsymbol{\sigma}^\delta \in \mathbb{R}^{D_g \times k}$.
\subsection{Uncertainty-Aware Multimodal Fusion (UAMF)}

Given the modality-wise variance vectors $\{\boldsymbol{\sigma}^\delta\}$, we first quantify the modality confidence based on relative uncertainty ratios:
\begin{equation}
\bm{C}^\delta_{\text{raw}} = {\bm1}-\frac{\boldsymbol{\sigma}^\delta+ \boldsymbol{\epsilon}}{\boldsymbol{\sigma}^a + \boldsymbol{\sigma}^v + \boldsymbol{\sigma}^t + \boldsymbol{\epsilon}}, 
\text{ } \delta \in \{t, a, v\},
\end{equation}
where $\boldsymbol{\epsilon}$ is a small constant for numerical stability, and the operations are performed element-wise along the feature dimension. This formulation assigns higher confidence to modalities with lower relative uncertainty. Softmax normalization is applied along the modality dimension to ensure fair competition:
\begin{equation}
\bm{C}^\delta = \text{softmax}(\bm{C}^\delta_{\text{raw}}) \in \mathbb{R}^{D_g \times k}.
\end{equation}
Then, the confidence-weighted features are obtained by elementwise multiplication with the original mean features:
\begin{equation}
\bm{R}^\delta_u = \bm{C}^\delta \odot \boldsymbol{\mu}^\delta, 
\text{ } \delta \in \{t, a, v\},
\end{equation}
where $\odot$ denotes element-wise multiplication, enabling automatic suppression of high-uncertainty features.

To model inter-utterance contextual dependencies, we employ a multi-head attention mechanism. The query, key, and value matrices are constructed as:
\begin{equation}
\begin{split}
    \bm{Q} &= \bm{W}^Q \operatorname{Concat}(\boldsymbol{\mu}^t, \boldsymbol{\mu}^a, \boldsymbol{\mu}^v)\in \mathbb{R}^{D_C \times k},\\
    \bm{K} &= \operatorname{Concat}(\bm{R}^t_u, \bm{R}^a_u, \bm{R}^v_u)\in \mathbb{R}^{D_C \times k}, \\
    \bm{V} &= \bm{W}^V \operatorname{Concat}(\boldsymbol{\mu}^t, \boldsymbol{\mu}^a, \boldsymbol{\mu}^v)\in \mathbb{R}^{D_C \times k},
\end{split}
\end{equation}
where $\bm{W}^Q$ and $\bm{W}^V$ are learnable projections and $D_C = 3D_g$. The uncertainty-aware attention output is computed as:
\begin{equation}
    \bm{A} = \text{softmax}\left(\frac{\bm{Q}^\top \bm{K}}{\sqrt{D_C}}\right)\bm{V} \in \mathbb{R}^{D_C \times k},
\end{equation}
allowing high-confidence modalities to receive larger attention weights.
The attended representations are further refined by a Transformer encoder to capture global conversational context:
\begin{equation}
\bm{H} = [\bm{h}_1,\ldots,\bm{h}_k] = \operatorname{TransEncoder}(\bm{A}),
\end{equation}
a residual connection is applied to preserve the original multimodal information and improve training stability:
\begin{equation}
\bm{F} = [\bm{f}_1,\ldots,\bm{f}_k] = \bm{H} + \operatorname{Concat}(\boldsymbol{\mu}^t, \boldsymbol{\mu}^a, \boldsymbol{\mu}^v),
\end{equation}
where $\bm{f}_i \in \mathbb{R}^{C}$ denotes the fused feature of utterance $u_i$. Each fused embedding $\bm{f}_i$ is fed into a classifier to obtain emotion probabilities $\bm{p}_i$, optimized using the standard cross-entropy loss $\mathcal{L}_{CE}$.
\subsection{Explicitly Supervised Loss (ESL)}
\begin{figure}[!htbp]
\centering
\vspace{-1.2em}
\includegraphics[width=0.48\textwidth]{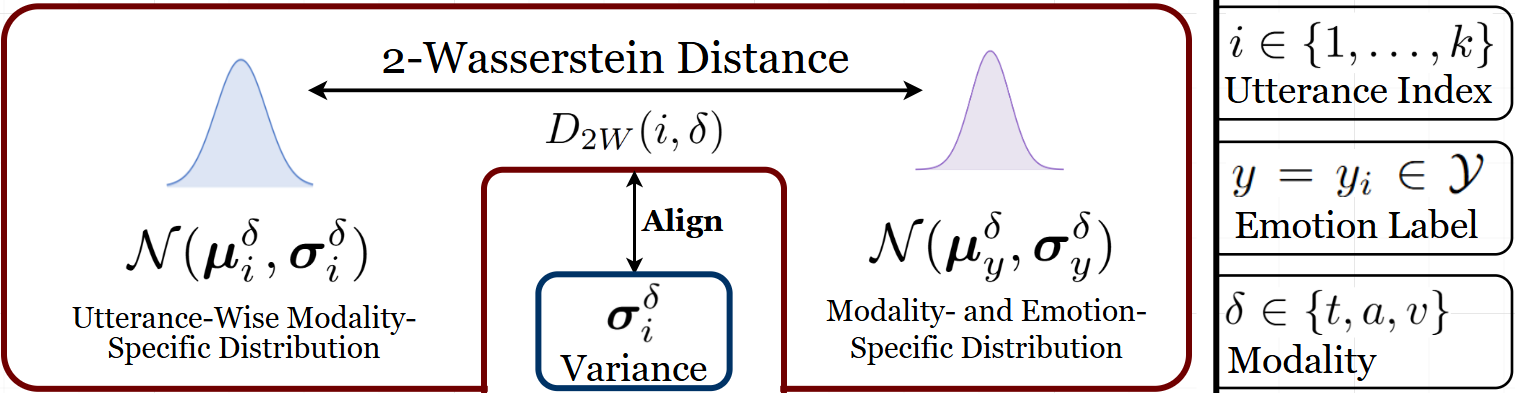}
\caption{Framework of explicitly supervised loss $\mathcal{L}_{ESL}^\delta$. ESL aims to align the predicted variance $\boldsymbol{\sigma}_i^\delta$ of each utterance with the 2-Wasserstein distance $D_{2W}(i, \delta)$ between the utterance-wise distribution and the corresponding global emotion center, providing explicit supervision for uncertainty estimation.}
\label{esl}
\vspace{-0.5em}
\end{figure}
We hypothesize that implicit uncertainty modeling, based solely on the cross-entropy loss $\mathcal{L}_{CE}$, is insufficient to accurately capture utterance-level uncertainty in multimodal emotion recognition. To address this limitation, we propose a novel explicitly supervised loss $\mathcal{L}_{ESL}$, as shown in Fig. 3, which directly supervises the variance outputs from the ContextDEM modules, explicitly encouraging the model to learn accurate and well-calibrated uncertainty representations across modalities.

\textbf{Alignment with Global Distributional Cluster Centers:}
To maintain consistency with the evolving feature representations, global cluster centers are updated once per training epoch using the current feature statistics and are not involved in gradient
backpropagation.
Unlike conventional methods that only store cluster means, we adopt a distributional representation of cluster centers to better capture the statistical properties.
For each emotion class $y$ and modality $\delta$, we compute both the mean and variance of the cluster center:

\begin{equation}
\begin{split}
    &\boldsymbol{\mu}_{y}^\delta = \frac{1}{|\mathcal{I}_{y}^\delta|} \sum_{{i_y} \in \mathcal{I}_{y}^\delta} \boldsymbol{\mu}_{i_y}^\delta \in \mathbb{R}^{D_g}, \text{ } \delta \in \{t, a, v\}, \\
    &\boldsymbol{\sigma}_{y}^\delta = \frac{1}{|\mathcal{I}_{y}^\delta|} \sum_{{i_y} \in \mathcal{I}_{y}^\delta} \boldsymbol{\sigma}_{i_y}^\delta \in \mathbb{R}^{D_g}, \text{ } \delta \in \{t, a, v\},
\end{split}
\end{equation}
where $\mathcal{I}_{y}^\delta = \{i_y \mid y_{i_y} = y \in \cal Y\}$ denotes the set of utterance indices with emotion label $y$, $D_g$ is the feature dimension, $\boldsymbol{\mu}_i^\delta$ and $\boldsymbol{\sigma}_i^\delta$ are the mean and variance vectors obtained from the ContextDEM modules, and $|\mathcal{I}_{y}^\delta|$ is the number of utterances with emotion class $y$.
For a dataset with 3 modalities and 6 emotion labels, 18 global distributional cluster centers are maintained.

To measure the deviation between an utterance's distributional representation and its emotion cluster center, we employ the 2-Wasserstein (2W) distance:

\begin{equation}
    D_{2W}(i, \delta) =
    \lVert\boldsymbol{\mu}_i^\delta - \boldsymbol{\mu}_{y_i}^\delta\rVert_2^2
    +
    \lVert\boldsymbol{\sigma}_i^\delta - \boldsymbol{\sigma}_{y_i}^\delta\rVert_2^2,
\end{equation}
where $y_i$ is the emotion label of utterance $i$, and
$\boldsymbol{\mu}_{y_i}^\delta$ and $\boldsymbol{\sigma}_{y_i}^\delta$ denote the mean and variance vectors of the cluster center corresponding to emotion $y_i \in \cal Y$ and modality $\delta$. The 2-Wasserstein distance can naturally account for the distributional uncertainty.
The explicit supervision loss is then formulated as:
\begin{equation}
    \mathcal{L}_{ESL}^\delta = \frac{1}{k} \sum_{i=1}^{k} \left\lVert \alpha \sqrt{\boldsymbol{\sigma}_i^\delta} - \sqrt{D_{2W}(i, \delta)} \right\rVert_2^2,
\end{equation}
where $k$ is the number of utterances in the conversation, and $\alpha$ is a learnable uncertainty scaling factor that addresses potential scale mismatches between the predicted variance and the embedding deviations.
\subsection{Jointly Training}
The final training objective combines the classification loss and the explicitly supervised uncertainty loss for all three modalities:
\begin{equation}
    \mathcal{L} = \mathcal{L}_{CE} + \kappa (\mathcal{L}_{ESL}^t + \mathcal{L}_{ESL}^a + \mathcal{L}_{ESL}^v),
\end{equation}
where $\kappa$ denotes the weighting coefficient for the explicitly supervised uncertainty loss. To stabilize training, the explicitly supervised loss $\mathcal{L}_{ESL}$ is introduced after an initial training phase. Specifically, the weight $\kappa$ remains 0 before a predefined start epoch $ep_{\text{start}}$, and is then set to a fixed constant $\lambda$:

\begin{equation}
    \kappa(\text{epoch}) =
    \begin{cases}
        0, & \text{if epoch} < ep_{\text{start}}, \\
        \lambda , & \text{otherwise}.
    \end{cases}
\end{equation}

\section{Experimental Settings}

\subsection{Datasets and Evaluation Metric}

\textbf{IEMOCAP} \cite{busso2008iemocap} is a widely used dataset for emotion recognition in conversation. For partitioning the data, we utilized the commonly used “Leave-One-Session-Out” (LOSO) strategy. 
\textbf{MELD} \cite{poria2018meld} is a multi-modal, multi-speaker conversational dataset derived from the TV series “Friends”. To ensure a fair comparison, we followed the predefined train/val/test splits provided by the dataset, ensuring data allocation aligned with \cite{hu2022mm}. 
Similar to other recent studies, we adopt the accuracy and weighted F1-score (w-F1) for evaluation. 

\subsection{Implementation Details}
For the feature extraction, we used RoBERTa \cite{liu2019roberta} for text, Wav2vec2.0 \cite{baevski2020wav2vec} for audio, and CLIP \cite{radford2021learning} for visual modality. All experiments were conducted on a single NVIDIA RTX 4090 GPU, trained for 40 and 50 epochs on IEMOCAP and MELD with a batch size of 15 and 100, respectively. The model was optimized using the Adam optimizer with a learning rate of $2 \times 10^{-4}$, loss weight of $\lambda = 2 \times 10^{-5}$ with the starting epoch $ep_{\text{start}}=10$, and a dropout rate of 0.2. Finally, we averaged the weights from the last 10 checkpoints to obtain the final model for evaluation.

\section{Experimental Results and Analysis}

\subsection{Effect of Different Fusion Methods and Overall Results} 
\begin{table*}[htbp]
\caption{Accuracy and weighted-average F1 score (w-F1) compared with other baselines. 
Best and second-best results are displayed in \textbf{bold} and \underline{underline}, respectively.}
\vspace{-0.6em}
    \centering
    \renewcommand{\arraystretch}{1.05}
    \begin{tabular}{c|c|c|c|c|c|c|c|c|c|c|c}
    \hline
    \multirow{2}{*}{\textbf{Baseline Model}} & \textbf{Proposed} & \multicolumn{8}{c|}{\textbf{IEMOCAP}} & \multicolumn{2}{c}{\textbf{MELD}} \\ \cline{3-12} 
    & \textbf{Year} & \textit{Happy} & \textit{Sad} & \textit{Neutral} & \textit{Angry} & \textit{Excited} & \textit{Frustrated} & \textbf{Acc} & {\textbf{w-F1}} & \textbf{Acc} & {\textbf{w-F1}} \\ \hline
    DialogRNN \cite{majumder2019dialoguernn} & 2019 & 32.20 & 80.26 & 57.89 & 62.82 & 73.87 & 59.76 & 63.52 & 62.89 & 60.31 & 57.66 \\
    DialogGCN \cite{ghosal2019dialoguegcn} & 2019 & 42.75 & 84.54 & 63.54 & 64.19 & 63.08 &  66.99 & 65.25 & 64.18 & - & 58.10 \\
    MMGCN \cite{hu2021mmgcn} & 2021 & 45.14 & 77.16 & 64.36 & 68.82 & 74.71 & 61.40 & 66.36 & 66.26 & 60.42 & 58.31 \\
    M2FNet \cite{chudasama2022m2fnet} & 2022 & - & - & - & - & - & - & 69.69 & 69.86 & \underline{67.85} & 66.71 \\
    CFN-ESA \cite{li2023cfn} & 2023 & 53.67 & 80.60 & \underline{71.65} & 70.32 & 74.82 & 68.06 & 71.04 & 70.78 & \underline{67.85} & 66.70 \\
    AdaIGN \cite{AdaIGN} & 2024 & 53.04 & 81.47 & 71.26 & 65.87 & 76.34 & 67.79 & - & 70.74 & - & \underline{66.79} \\
    MDAG \cite{curriculum} & 2024 & 45.26 & \underline{81.40} & 69.53 & 70.33 & 71.61 & 66.94 & 69.11 & 69.08 & 64.41 & 64.00 \\
    DER-GCN \cite{DER-GCN}  & 2025 & 58.80 & 79.80 & 61.50 & \textbf{72.10} & 73.30 & 67.80 & 69.70 & 69.40 & 66.80 & 66.10 \\
   FEMI \cite{FEMI} & 2025 & 60.60 & \textbf{85.55} & 70.92 & 70.98 & 75.13 & \underline{69.00} & 71.97 & \underline{73.53} & 64.88 & 66.41 \\
    \hline \hline 
            \textbf{EmoEUS} & \textbf{Ours} & \textbf{77.42} & 72.87 & \textbf{72.04} & \underline{71.20} & \textbf{78.66} & \textbf{73.58} & \textbf{74.33} & \textbf{74.36} & \textbf{68.32} & \textbf{67.53} \\
    \hline
    \end{tabular}
    \label{tab:results}
    \vspace{-1.27em}
\end{table*}
Table~\ref{tab:results} shows the detailed results on both datasets. The proposed EmoEUS performs the best among all models, demonstrating its effectiveness. By explicitly modeling uncertainty through $\mathcal{L}_{ESL}$, EmoEUS effectively identifies and emphasizes reliable modalities while suppressing uncertain ones, leading to more robust multimodal representations. 

While EmoEUS shows slightly lower scores on the \textit{Sad} and \textit{Angry} classes compared to some baselines, it exhibits significantly more balanced and robust performance across all emotion categories (achieving w-F1 scores above 70\% for every emotion class on IEMOCAP), avoiding the severe bias towards specific classes observed in other methods and thereby leading to superior overall accuracy and w-F1 scores. 
\begin{table}[ht]
\vspace{-0.15em}
    \caption{Performance of using different model settings.}
    \vspace{-0.2em}
    \centering
    \renewcommand{\arraystretch}{1.05}
    \begin{tabular}{c|c|c|c|c}
    \hline
 \multirow{2}{*}{\textbf{Fusion Method}} & \multicolumn{2}{c|}{\textbf{IEMOCAP}} & \multicolumn{2}{c}{\textbf{MELD}} \\ \cline{2-5} 
 &\textbf{Acc}& \textbf{w-F1} &\textbf{Acc} & \textbf{w-F1}\\ \hline
Concat & 71.90 & 72.20 & 66.82 & 65.57 \\
Attention & 72.39 & 72.33 & 66.64 & 65.75 \\
TransFormer & 72.53 & 72.57 & 67.64 & 66.58 \\ \hline \hline
 \textbf{EmoEUS (Full)} & \textbf{74.33} & \textbf{74.36} & \textbf{68.32} & \textbf{67.53}  \\ 
 w/o \textbf{$\mathcal{L}_{ESL}$} & 73.32 & 73.33 & 67.78 & 66.73 \\
 w/o UAMF& 72.63 & 72.69 & 66.53 & 66.01 \\ \hline 

    \end{tabular}
    \label{tab3}
        \vspace{-1.5em}
\end{table}

Table~\ref{tab3} presents the performance of different fusion methods. “Concat” directly concatenates features ($\bm{R}^\delta$) from different modalities and feeds them to the classifier. “Attention” and “Transformer” employ a multi-head attention module and a simple Transformer encoder to fuse the concatenated multimodal features, respectively. Compared with these baselines, EmoEUS consistently achieves better performance on both datasets, demonstrating the effectiveness of the uncertainty-aware fusion mechanism.
Performance drops in the variants without UAMF and $\mathcal{L}_{ESL}$ further verify the importance of these components. Removing UAMF degrades performance, showing that uncertainty-aware fusion is crucial for multimodal integration. The variant without $\mathcal{L}_{ESL}$ also declines, suggesting that explicit uncertainty supervision helps learn more accurate and well-calibrated uncertainty estimates, which further facilitate more effective uncertainty-aware fusion.
        \vspace{-0.9em}

\subsection{Performance on Ambiguous Emotion Pairs}
The results in Table~\ref{tab:misclass} demonstrate that the proposed EmoEUS effectively reduces misclassification rates between ambiguous emotion pairs (e.g., \textit{Hap-Exc} denotes misclassifying ``Happy'' as ``Excited''), thereby improving overall emotion prediction performance. Notably, the full model with explicit uncertainty supervision $\mathcal{L}_{ESL}$ achieves further reductions in misclassification rates for most ambiguous pairs, which enables the model to learn more separable and discriminative representations for semantically similar emotions. By directly supervising the variance outputs and aligning them with emotion-specific distributional cluster centers, the model better captures fine-grained uncertainty patterns, thereby leading to more robust overall prediction performance.
\begin{table}[!htbp]
    \setlength{\tabcolsep} {0.5mm}
    \renewcommand{\arraystretch}{1.05}
    \caption{Misclassification Rates Reduction on the Ambiguous Emotion Pairs in IEMOCAP (\textbf{Lower is Better}).}
            \vspace{-0.5em}
    \centering
    \begin{tabular}{c|c|c|c|c|c}
    \hline
    {\textbf{Methods}} & \textit{Hap-Neu} & \textit{Hap-Exc} & \textit{Neu-Exi} & \textit{Sad-Fru} & \textit{Exc-Hap} \\ \hline
    TransFormer & 7.68 & 16.74 & 5.41 & 24.62 & 7.54 \\ \hline \hline
    w/o $\mathcal{L}_{ESL}$$^\mathrm{*}$ & 7.57 & 19.05 & 5.23 & 16.92 & 6.56  \\
    \textbf{EmoEUS (Full)} & \textbf{6.51} & \textbf{14.88} & \textbf{4.96} & \textbf{16.92} & \textbf{6.89} \\  \hline
  \multicolumn{6}{l}{$^{\mathrm{*}}$Train without the explicit uncertainty supervision loss $\mathcal{L}_{ESL}$} 
    \end{tabular}
    \label{tab:misclass}
   \vspace{-2.4em}
\end{table}
        \vspace{-0.5em}
\subsection{Ablation Studies on ContextDEM}
Table~\ref{tab:residual} demonstrates the advantages of distributional representation over point estimation. 
Modeling both mean ($\boldsymbol{\mu}^\delta$) and variance ($\boldsymbol{\sigma}^\delta$) with the 2-Wasserstein distance consistently outperforms single-point representation with the standard MSE distance across all settings, validating that explicitly capturing uncertainty through variance leads to more robust feature learning. Furthermore, the residual connection proves effective in preserving the original contextual features while integrating the global interactions from the Transformer encoder.
\begin{table}[!ht]
\renewcommand{\arraystretch}{1.15}

\centering
 \vspace{-0.8em}
    \caption{Effect of Distributional Representation Modeling and Residual Connection on ContextDEM. For comparison, we employed an MSE distance for point representation using $(\boldsymbol{\mu}^\delta)$.}
    \begin{tabular}{c|c|c|c|c|c}
    \hline
    \multirow{2}{*}{} & \textbf{Residual} & \multicolumn{2}{c|}{\textbf{IEMOCAP}} & \multicolumn{2}{c}{\textbf{MELD}} \\ \cline{3-6} 
     &\textbf{Connection}& \textbf{Acc} & \textbf{w-F1} & \textbf{Acc} &  \textbf{w-F1} \\ \hline
    \multicolumn{6}{c}{Point-Rep ($\boldsymbol{\mu}^\delta$) with MSE Distance $D_{MSE}$} \\ \hline
     ($\boldsymbol{\mu}^\delta$)  & \textbf{×} & 72.86 & 72.42 & 67.53 & 67.45 \\
     ($\boldsymbol{\mu}^\delta$) & \textbf{$\checkmark$} & 73.28 & 72.47 & 66.98 & 67.26 \\ \hline
    \multicolumn{6}{c}{Distribution-Rep ($\boldsymbol{\mu}^\delta, \boldsymbol{\sigma}^\delta$) with 2-Wasserstein Distance $D_{2W}$} \\ \hline
     ($\boldsymbol{\mu}^\delta, \boldsymbol{\sigma}^\delta$) & \textbf{×} &73.98 & 73.89 & 68.08 & 67.31 \\
   ($\boldsymbol{\mu}^\delta, \boldsymbol{\sigma}^\delta$) & \textbf{$\checkmark$} & \textbf{74.33} & \textbf{74.36} & \textbf{68.32} & \textbf{67.53} \\ \hline
    \end{tabular}
\label{tab:residual}
 \vspace{-\baselineskip}
 \vspace{-1.2em}
\end{table}

\section{Conclusions and Future Work} \label{conclusions}
We propose the EmoEUS, a novel framework that incorporates explicit uncertainty supervision for MERC. Our approach introduces a context-level distribution estimator module (ContextDEM) to model utterance-level distribution, dynamically weights modality reliability through an uncertainty-aware multimodal fusion (UAMF) mechanism, and supervises the uncertainty estimation using an explicitly supervised loss that aligns predicted variances with distributional deviations. Extensive experiments on IEMOCAP and MELD demonstrate state-of-the-art performance, with significant reductions in misclassification rates between ambiguous emotion pairs, validating the effectiveness of explicit uncertainty supervision. In future work, we will enhance the uncertainty estimation module to better capture cross-speaker emotional dynamics in real-time settings.
\section{Acknowledgment} 
\label{Acknowledgment}
The work presented in this article is supported by the Research Platform for Advanced Audio and Speech Signal Processing (P0049192) funded by Innovation Technology Co. Ltd.

\section{Use of Generative AI Disclosure}
\label{GenAI}
Generative AI tools were used only for language polishing and formatting assistance. All scientific content, experiments and analyses were produced and verified by the authors. 

\bibliographystyle{IEEEtran}
\bibliography{mybib}

@article{poria2019emotion,
  title={Emotion recognition in conversation: Research challenges, datasets, and recent advances},
  author={Poria, Soujanya and Majumder, Navonil and Mihalcea, Rada and Hovy, Eduard},
  journal={IEEE Access},
  volume={7},
  pages={100943--100953},
  year={2019},
  publisher={IEEE}
}

@inproceedings{mm-nodeformer,
   author = {Huang, Zilong and Mak, Man-Wai and Lee, Kong Aik},
   title = {{MM-NodeFormer}: Node Transformer Multimodal Fusion for Emotion Recognition in Conversation},
   booktitle = {Proc. Interspeech},
   pages = {4069-4073},
   type = {Conference Proceedings},
   year={2024}
}

@article{ghosal2019dialoguegcn,
  title={{DialogueGCN}: A graph convolutional neural network for emotion recognition in conversation},
  author={Ghosal, Deepanway and Majumder, Navonil and Poria, Soujanya and Chhaya, Niyati and Gelbukh, Alexander},
  journal={arXiv preprint arXiv:1908.11540},
  year={2019}
}

@article{wu2014survey,
  title={Survey on audiovisual emotion recognition: {Databases}, features, and data fusion strategies},
  author={Wu, Chung-Hsien and Lin, Jen-Chun and Wei, Wen-Li},
  journal={APSIPA Transactions on Signal and Information Processing},
  volume={3},
  pages={e12},
  year={2014},
  publisher={Cambridge University Press}
}

@inproceedings{hu2022mm,
  title={{MM-DFN}: Multimodal dynamic fusion network for emotion recognition in conversations},
  author={Hu, Dou and Hou, Xiaolong and Wei, Lingwei and Jiang, Lianxin and Mo, Yang},
  booktitle={Proc. IEEE International Conference on Acoustics, Speech and Signal Processing},
  pages={7037--7041},
  year={2022}
}

@article{hu2021mmgcn,
  title={{MMGCN}: Multimodal fusion via deep graph convolution network for emotion recognition in conversation},
  author={Hu, Jingwen and Liu, Yuchen and Zhao, Jinming and Jin, Qin},
  journal={arXiv preprint arXiv:2107.06779},
  year={2021}
}

@article{busso2008iemocap,
  title={{IEMOCAP}: Interactive emotional dyadic motion capture database},
  author={Busso, Carlos and Bulut, Murtaza and Lee, Chi-Chun and Kazemzadeh, Abe and Mower, Emily and Kim, Samuel and Chang, Jeannette N and Lee, Sungbok and Narayanan, Shrikanth S},
  journal={Language Resources and Evaluation},
  volume={42},
  pages={335--359},
  year={2008},
  publisher={Springer}
}

@article{poria2018meld,
  title={{MELD}: A multimodal multi-party dataset for emotion recognition in conversations},
  author={Poria, Soujanya and Hazarika, Devamanyu and Majumder, Navonil and Naik, Gautam and Cambria, Erik and Mihalcea, Rada},
  journal={arXiv preprint arXiv:1810.02508},
  year={2018}
}

@inproceedings{glodek2011multiple,
  title={Multiple classifier systems for the classification of audio-visual emotional states},
  author={Glodek, Michael and Tschechne, Stephan and Layher, Georg and Schels, Martin and Brosch, Tobias and Scherer, Stefan and K{\"a}chele, Markus and Schmidt, Miriam and Neumann, Heiko and Palm, G{\"u}nther and others},
  booktitle={Affective Computing and Intelligent Interaction: Fourth International Conference, ACII 2011, Memphis, TN, USA, October 9--12, 2011, Proceedings, Part II},
  pages={359--368},
  year={2011},
  organization={Springer}
}

@inproceedings{majumder2019dialoguernn,
  title={{DialogueRNN}: An attentive rnn for emotion detection in conversations},
  author={Majumder, Navonil and Poria, Soujanya and Hazarika, Devamanyu and Mihalcea, Rada and Gelbukh, Alexander and Cambria, Erik},
  booktitle={Proceedings of the AAAI Conference on Artificial Intelligence},
  volume={33},
  pages={6818--6825},
  year={2019}
}

@article{liu2019roberta,
  title={{RoBERTa}: A robustly optimized BERT pretraining approach},
  author={Liu, Yinhan and Ott, Myle and Goyal, Naman and Du, Jingfei and Joshi, Mandar and Chen, Danqi and Levy, Omer and Lewis, Mike and Zettlemoyer, Luke and Stoyanov, Veselin},
  journal={arXiv preprint arXiv:1907.11692},
  year={2019}
}

@article{baevski2020wav2vec,
  title={wav2vec 2.0: A framework for self-supervised learning of speech representations},
  author={Baevski, Alexei and Zhou, Yuhao and Mohamed, Abdelrahman and Auli, Michael},
  journal={Advances in Neural Information Processing Systems},
  volume={33},
  pages={12449--12460},
  year={2020}
}

@inproceedings{radford2021learning,
  title={Learning transferable visual models from natural language supervision},
  author={Radford, Alec and Kim, Jong Wook and Hallacy, Chris and Ramesh, Aditya and Goh, Gabriel and Agarwal, Sandhini and Sastry, Girish and Askell, Amanda and Mishkin, Pamela and Clark, Jack and others},
  booktitle={Proc. International Conference on Machine Learning},
  pages={8748--8763},
  year={2021}
}

@inproceedings{chudasama2022m2fnet,
  title={{M2FNET}: Multi-modal fusion network for emotion recognition in conversation},
  author={Chudasama, Vishal and Kar, Purbayan and Gudmalwar, Ashish and Shah, Nirmesh and Wasnik, Pankaj and Onoe, Naoyuki},
  booktitle={Proceedings of the IEEE/CVF Conference on Computer Vision and Pattern Recognition},
  pages={4652--4661},
  year={2022}
}

@inproceedings{shi-huang-2023-multiemo,
    title = "{M}ulti{EMO}: An Attention-Based Correlation-Aware Multimodal Fusion Framework for Emotion Recognition in Conversations",
    author = "Shi, Tao  and
      Huang, Shao-Lun",
    editor = "Rogers, Anna  and
      Boyd-Graber, Jordan  and
      Okazaki, Naoaki",
    booktitle = "Proceedings of the 61st Annual Meeting of the Association for Computational Linguistics (Volume 1: Long Papers)",
    month = jul,
    year = "2023",
    address = "Toronto, Canada",
    publisher = "Association for Computational Linguistics",
    url = "https://aclanthology.org/2023.acl-long.824/",
    doi = "10.18653/v1/2023.acl-long.824",
    pages = "14752--14766",
}

@INPROCEEDINGS{MAP,
  author={Ji, Yatai and Wang, Junjie and Gong, Yuan and Zhang, Lin and Zhu, Yanru and Wang, Hongfa and Zhang, Jiaxing and Sakai, Tetsuya and Yang, Yujiu},
  booktitle={2023 IEEE/CVF Conference on Computer Vision and Pattern Recognition (CVPR)}, 
  title={MAP: Multimodal Uncertainty-Aware Vision-Language Pre-training Model}, 
  year={2023},
  volume={},
  number={},
  pages={23262-23271},
  doi={10.1109/CVPR52729.2023.02228}}

@article{li2023cfn,
  title={{CFN-ESA}: A Cross-Modal Fusion Network with Emotion-Shift Awareness for Dialogue Emotion Recognition},
  author={Li, Jiang and Liu, Yingjian and Wang, Xiaoping and Zeng, Zhigang},
  journal={arXiv preprint arXiv:2307.15432},
  year={2023}
}

@ARTICLE{DER-GCN,
  author={Ai, Wei and Shou, Yuntao and Meng, Tao and Li, Keqin},
  journal={IEEE Transactions on Neural Networks and Learning Systems}, 
  title={{DER-GCN}: Dialog and Event Relation-Aware Graph Convolutional Neural Network for Multimodal Dialog Emotion Recognition}, 
  year={2025},
  volume={36},
  number={3},
  pages={4908-4921},
  doi={10.1109/TNNLS.2024.3367940}}

@article{AdaIGN,
   author = {Tu, Geng and Xie, Tian and Liang, Bin and Wang, Hongpeng and Xu, Ruifeng},
   title = {Adaptive Graph Learning for Multimodal Conversational Emotion Detection},
   journal = {Proceedings of the AAAI Conference on Artificial Intelligence},
   volume = {38},
   number = {17},
   ISSN = {2374-3468},
   DOI = {10.1609/aaai.v38i17.29876},
   url = {https://ojs.aaai.org/index.php/AAAI/article/view/29876},
   year = {2024},
   type = {Journal Article}
}

@INPROCEEDINGS{uncertain1,
  author={Wang, Xuechen and Zhao, Shiwan and Sun, Haoqin and Wang, Hui and Zhou, Jiaming and Qin, Yong},
  booktitle={ICASSP 2025 - 2025 IEEE International Conference on Acoustics, Speech and Signal Processing (ICASSP)}, 
  title={Enhancing Multimodal Emotion Recognition through Multi-Granularity Cross-Modal Alignment}, 
  year={2025},
  volume={},
  number={},
  pages={1-5},
  doi={10.1109/ICASSP49660.2025.10889156}}

@INPROCEEDINGS{uncertain2,
  author={Sun, Haoqin and Zhao, Shiwan and Li, Shaokai and Kong, Xiangyu and Wang, Xuechen and Zhou, Jiaming and Kong, Aobo and Chen, Yong and Zeng, Wenjia and Qin, Yong},
  booktitle={ICASSP 2025 - 2025 IEEE International Conference on Acoustics, Speech and Signal Processing (ICASSP)}, 
  title={Enhancing Emotion Recognition in Incomplete Data: A Novel Cross-Modal Alignment, Reconstruction, and Refinement Framework}, 
  year={2025},
  volume={},
  number={},
  pages={1-5},
  doi={10.1109/ICASSP49660.2025.10889485}}

@article{meng2024masked,
  title={Masked graph learning with recurrent alignment for multimodal emotion recognition in conversation},
  author={Meng, Tao and Zhang, Fuchen and Shou, Yuntao and Shao, Hongen and Ai, Wei and Li, Keqin},
  journal={IEEE/ACM Transactions on Audio, Speech, and Language Processing},
  year={2024},
  publisher={IEEE}
}

@inproceedings{cv1,
  title={Probabilistic face embeddings},
  author={Shi, Yichun and Jain, Anil K},
  booktitle={Proceedings of the IEEE/CVF International Conference on Computer Vision},
  pages={6902--6911},
  year={2019}
}

@inproceedings{cv2,
  title={Data uncertainty learning in face recognition},
  author={Chang, Jie and Lan, Zhonghao and Cheng, Changmao and Wei, Yichen},
  booktitle={Proceedings of the IEEE/CVF cConference on Computer Vision and Pattern Recognition},
  pages={5710--5719},
  year={2020}
}

@inproceedings{curriculum,
  title={Curriculum learning meets directed acyclic graph for multimodal emotion recognition},
  author={Nguyen, Cao-Bach and Le, Duc-Trong and Ha, Quang Thuy and others},
  booktitle={Proceedings of the 2024 Joint International Conference on Computational Linguistics, Language Resources and Evaluation (LREC-COLING 2024)},
  pages={4259--4265},
  year={2024}
}

@article{FEMI,
  title={Feature-Enhanced Multimodal Interaction model for emotion recognition in conversation},
  author={Fu, Yanping and Yan, XiaoYuan and Chen, Wei and Zhang, Jun},
  journal={Knowledge-Based Systems},
  volume={309},
  pages={112876},
  year={2025},
  publisher={Elsevier}
}

@article{SDT,
  title={A transformer-based model with self-distillation for multimodal emotion recognition in conversations},
  author={Ma, Hui and Wang, Jian and Lin, Hongfei and Zhang, Bo and Zhang, Yijia and Xu, Bo},
  journal={IEEE Transactions on Multimedia},
  volume={26},
  pages={776--788},
  year={2023},
  publisher={IEEE}
}

@ARTICLE{COLD,
  author={Tellamekala, Mani Kumar and Amiriparian, Shahin and Schuller, Björn W. and André, Elisabeth and Giesbrecht, Timo and Valstar, Michel},
  journal={IEEE Transactions on Pattern Analysis and Machine Intelligence}, 
  title={{COLD} {F}usion: Calibrated and Ordinal Latent Distribution Fusion for Uncertainty-Aware Multimodal Emotion Recognition}, 
  year={2024},
  volume={46},
  number={2},
  pages={805-822},
  doi={10.1109/TPAMI.2023.3325770}}

@article{li2025disentangling,
  title={Disentangling Speech Representations Learning With Latent Diffusion for Speaker Verification},
  author={Li, Zhe and Mak, Man-Wai and Chien, Jen-Tzung and Pilanci, Mert and Jin, Zezhong and Meng, Helen},
  journal={IEEE Transactions on Audio, Speech and Language Processing},volume={33},
  pages={3896-3907},  
  year={2025},
  publisher={IEEE}
}

@INPROCEEDINGS{extradis,
  author={Jin, Zezhong and Liu, Shujie and Li, Zhe and Gan, Chong-Xin and Huang, Zilong and Mak, Man-Wai and Lee, Kong Aik},
  booktitle={ICASSP 2026 - 2026 IEEE International Conference on Acoustics, Speech and Signal Processing (ICASSP)}, 
  title={Distilling Attention Knowledge for Speaker Verification}, 
  year={2026},
  volume={},
  number={},
  pages={16447-16451},
  doi={10.1109/ICASSP55912.2026.11464971}}

@inproceedings{extraidir,
  title     = {{IDIR: Identifying and Distilling Informative Relations for Speaker Verification}},
  author    = {Chong-Xin Gan and Zhe Li and Zezhong Jin and Zilong Huang and Man-Wai Mak and Kong Aik Lee},
  year      = {2025},
  booktitle = {{Interspeech 2025}},
  pages     = {5758--5762},
  doi       = {10.21437/Interspeech.2025-736},
  issn      = {2958-1796},
}

@inproceedings{extraren2025bav,
  title={{BAV-M}oss{F}ormer2: Enhanced {M}oss{F}ormer2 for Binaural Audio-Visual Speech Enhancement},
  author={Ren, Wenze and Li, Kai and Chao, Rong and Li, Junjie and Huang, Zilong and Ahmed, Shafique and Li, You-Jin and Hung, Kuo-Hsuan and Wang, Syu-Siang and Wang, Hsin-Min and others},
  booktitle={Proc. AVSEC 2025},
  pages={79--80},
  year={2025}
}

@article{extraunet,
  title={U{N}et-Based Fusion and Exponential Moving Average Adaptation for Noise-Robust Speaker Recognition},
  author={Gan, Chong-Xin and Bell, Peter and Mak, Man-Wai and Li, Zhe and Jin, Zezhong and Huang, Zilong and Lee, Kong Aik},
  journal={arXiv preprint arXiv:2604.25624},
  year={2026}
}

@INPROCEEDINGS{extradenoising,
  author={Jin, Zezhong and Tu, Youzhi and Li, Zhe and Huang, Zilong and Gan, Chong-Xin and Mak, Man-Wai},
  booktitle={ICASSP 2025 - 2025 IEEE International Conference on Acoustics, Speech and Signal Processing (ICASSP)}, 
  title={Denoising Student Features with Diffusion Models for Knowledge Distillation in Speaker Verification}, 
  year={2025},
  volume={},
  number={},
  pages={1-5},
  doi={10.1109/ICASSP49660.2025.10889980}}

@inproceedings{disstilmerc,
  title={Unimodal-driven distillation in multimodal emotion recognition with dynamic fusion},
  author={Li, Jiagen and Yu, Rui and Huang, Huihao and Yan, Huaicheng},
  booktitle={2025 IEEE International Conference on Multimedia and Expo (ICME)},
  pages={1--6},
  year={2025},
  organization={IEEE}
}

\end{document}